\documentclass{article}
\usepackage{ismir,amsmath,cite,url}
\usepackage{graphicx}
\usepackage{color}
\usepackage{booktabs}
\usepackage{makecell}
\usepackage{refcount}

\title{Towards an efficient deep learning model for musical onset detection}

\twoauthors
  {Rong Gong} {Music Technology Group\\Universitat Pompeu Fabra, Barcelona \\ {\tt rong.gong@upf.edu}}
  {Xavier Serra} {Music Technology Group\\Universitat Pompeu Fabra, Barcelona\\ {\tt xavier.serra@upf.edu}}

\begin{document}

\maketitle
\begin{abstract}
In this paper, we propose an efficient and reproducible deep learning model for musical onset detection (MOD). We first review the state-of-the-art deep learning models for MOD, and identify their shortcomings and challenges: (i) the lack of hyper-parameter tuning details, (ii) the non-availability of code for training models on other datasets, and (iii) ignoring the network capability when comparing different architectures. Taking the above issues into account, we experiment with seven deep learning architectures. The most efficient one achieves equivalent performance to our implementation of the state-of-the-art architecture. However, it has only 28.3\% of the total number of trainable parameters compared to the state-of-the-art. Our experiments are conducted using two different datasets: one mainly consists of instrumental music excerpts, and another developed by ourselves includes only solo singing voice excerpts. Further, inter-dataset transfer learning experiments are conducted. The results show that the model pre-trained on one dataset fails to detect onsets on another dataset, which denotes the importance of providing the implementation code to enable re-training the model for a different dataset. Datasets, code and a Jupyter notebook running on Google Colab are publicly available to make this research understandable and easy to reproduce.

\end{abstract}
\section{Introduction}\label{sec:introduction}
Musical onset detection (MOD) is a prerequisite step for many MIR tasks, such as beat tracking, e.g.\cite{Bock2014AMA}, tempo estimation, e.g.\cite{bock2015accurate}, drum transcription, e.g.\cite{Vogl2017DrumTV}, note transcription, e.g.\cite{bock2012polyphonic} and singing voice syllable onset detection, e.g.\cite{gong2017score,pons2017score}. MOD performance has been boosted recently by deep learning, which led to the best onset detection accuracy in MIREX contest\footnote{\url{http://nema.lis.illinois.edu/nema_out/mirex2017/results/aod/}}\cite{schluter2014improved}. Hutson\cite{Hutson2018} revealed that some recent machine learning studies face reproducibility problems -- algorithm's code or dataset is often not made available, algorithm implementation details are not sufficient to reproduce the claimed results. Not coincidentally, ICML 2017 (International Conference on Machine Learning) included a ``Reproducibility in machine learning Workshop"\footnote{\url{https://goo.gl/jby4Hm}} to make researchers aware of the need for conducting reproducible research in machine learning field.

\subsection{Related work}\label{sec:related_work}

Most of the MOD methods follow this pipeline -- (1) calculating audio input representation, (2) onset detection function (ODF) computation, (3) onset selection.

Various audio input representations have been used for the first step of the pipeline, such as filtered logarithmic magnitude and phase spectrum\cite{Bello2003phase,bock2013local}. The former can be subdivided by the filterbank type -- Bark scale bands\cite{Bock2012OnlineRO}, Mel scale bands\cite{eyben2010universal,schluter2014improved} or constant-Q bands\cite{Lacoste2007supervised,Bock2012EvaluatingTO}.

Depending on the techniques used, we classify ODF computation methods into three categories:

\noindent\textbf{Unsupervised methods}: Earlier methods in this category are based on calculating temporal, spectral, phase, time-frequency or complex domain features, such as energy envelope, high-frequency content, spectral difference, phase deviation and negative log-likelihoods. Bello et al.\cite{bello2005tutorial} and Dixon\cite{dixon2006onset} both reviewed these methods thoroughly. The state-of-the-art methods in this category are based on spectral flux feature\cite{Bock2012EvaluatingTO}. Some variants such as \textit{SuperFlux}\cite{Bock2013MaximumFV}, local group delay weighting\cite{bock2013local} are proposed to suppress the negative effect of vibrato, primarily for pitched non-percussive instruments. The advantage of these methods is that no data is needed for training the model, and they are computationally efficient and can often operate in online real-time scenarios.

\noindent\textbf{Non-deep learning-based supervised methods}: Some methods in this category are based on probabilistic models, such as using Gaussian autoregressive models to detect the onset change point\cite{bello2005tutorial}. Toh et al.\cite{toh2008multiple} proposed a method using two Gaussian Mixture Models to classify audio features of onset frames and non-onset frames. Chen\cite{Chen2016Onset} detected the onset candidates from two ODFs, extracted features around these candidates, then used support vector machine technique to classify them.

\noindent\textbf{Deep learning-based supervised methods}: The state-of-the-art performance in the MIREX Audio Onset Detection is defined by deep learning-based methods. Eyben et al.\cite{eyben2010universal} proposed using recurrent neural networks (RNNs) with LSTM units to predict the input frames binarily as onset or non-onset. Schl\"{u}ter and B\"{o}ck\cite{schluter2014improved} used the similar idea but replaced RNNs by convolutional neural networks (CNNs) and achieved the best performance in the MIREX Audio Onset Detection task. Vogl et al.\cite{Vogl2017DrumTV} used convolutional-recurrent neural networks (CRNNs) to detect drum onset and produced a better score than CNNs on several percussion datasets.

The last step of the pipeline -- onset selection can be done by peak-picking\cite{Bock2012EvaluatingTO} or hidden Markov model (HMM) inference\cite{gong2017score,pons2017score} if the musical score is available.

In this research, we use the above three steps pipeline and focus on the deep learning-based supervised methods category. In the next section, we investigate the existing shortcomings and challenges in the previous deep learning-based works.

\subsection{Shortcomings and challenges}\label{sec:challenges}

First, some authors tended to not or very limitedly show the process that how they come to find the network architectures regarding the layer type and neuron numbers etc.\cite{eyben2010universal,schluter2014improved}. The hyperparameter tuning is an important engineering topic in machine learning, which searches a set of suitable hyperparameters for a learning algorithm and a particular dataset. If this process is not clarified, latecomer researchers might dissipate time on the tuning process which has been tested and proved to be not effective. 


Second, some authors didn't publicize the experiment code, which makes it hard to reproduce the claimed results by implementing the details described in the paper. Nowadays, researchers use various deep learning frameworks (Tensorflow, Pytorch, Keras, etc.) in their research. These frameworks usually have different default parameter settings or different algorithm implementations, and it is generally not possible to cover all the implementation details within a few pages. Thus, it is vital to provide the code for reproducing an exact experiment result. Further, while pre-trained models are available sometimes (e.g. \cite{madmom2016}), this is not enough, since a deep learning model performance heavily depends on the training dataset being used, and one might want to re-train the state-of-the-art model on a different dataset than the source dataset. In sections \ref{sec:transfer_learn} and \ref{sec:transfer_results}, we clarify this point by an inter-dataset experiment, in which the model pre-trained on one dataset fails on another.

Lastly, some authors didn't consider the network capabilities when comparing different network architectures. A network with more parameters has more capacity to `memorize' or overfit the training set than the one with fewer parameters, thus making the comparison unfair. 
In this research, we keep the total number of trainable parameters (TNoTP) almost equal when comparing different architectures. TNoTP is a metric for measuring the network capacity\cite{Collins2017capacity,Melis2017nlm}. 

\subsection{Contribution}\label{sec:contributions}

To find an efficient deep learning architecture for MOD, we experiment seven architectures on two different datasets (section \ref{sec:architectures}). We experiment two onset selection methods and shows the preferability of using the score-informed method if the musical score is available as a side information (section \ref{sec:nn_results}). We also conduct inter-dataset transfer learning experiments (section \ref{sec:transfer_learn}), of which the results stress the necessity of providing the model training code (section \ref{sec:transfer_results}). Lastly, to max out the reproducibility of this work, we make the datasets, experiment code publicly available, and simplify the process of using the code to re-train the model on user's dataset (section \ref{sec:reproducibility}).





\section{Datasets}\label{sec:datasets}
We use two datasets for model training and evaluation. The first dataset is used in B\"{o}ck et al.'s work\cite{Bock2012EvaluatingTO} and then used in several subsequent onset detection works, such as in \cite{Bock2013MaximumFV,schluter2014improved}  -- we call it \textbf{B\"{o}ck} dataset. It contains more than 25k onsets. Most content in this dataset is the complex mixture or solo instrumental excerpts. Only three excerpts are solo singing voice. We use the same 8-folds cross-validation configuration as in Schl\"{u}ter and B\"{o}ck's work\cite{schluter2014improved}. This dataset is available on request.
\begin{table}[ht]
	\centering
	\caption{Statistics of the jingju dataset. Phrases: singing phrases or melodic lines.}
	\label{table:detail_info_jingju_dataset}
	\begin{tabular}{l|ccc}
		\toprule
		& \#Recordings & \#Phrases & \#Syllables \\
		\midrule
		Train           & 85 & 883 & 8368  \\
		Test   			& 15 & 133 & 1203  \\
		\bottomrule
	\end{tabular}
\end{table}

The second dataset is a subset of a solo jingju singing voice dataset which has been jointly created by the researchers in (research institutes, omitted for the blind review)
-- we call it \textbf{jingju} dataset. 
It focuses on two most important jingju role-types (performing profile) \cite{repetto_creating_2014}: \textit{dan} (female) and \textit{laosheng} (old man). Jingju dataset contains 100 recordings, and manually annotated for each syllable onset in Praat\cite{Boersma2001} by the authors of this paper. The syllable onset detection evaluation is conducted on each musical phrase which has been pre-segmented manually. The statistics and train-test sets split are shown in table \ref{table:detail_info_jingju_dataset}. It is worth to mention that the artists, recording rooms and equipment used for the test set is completely different from the training set. This train-test split setup avoids the artist/room/equipment filtering effects in the evaluation process\cite{flexer2010effects}. The musical score is also included in this dataset, which provides the syllable duration prior information for the evaluation. This dataset is openly available\footnote{\url{https://goo.gl/qhG2xw}\label{fn:jingju_dataset}}

\section{Method}\label{sec:method}
\subsection{Audio input representation}\label{sec:preprocessing}
We use \textsc{Madmom}\cite{madmom2016} Python package to calculate the log-mel spectrogram of the student's singing audio. The frame size and hop size of the spectrogram are respectively 46.4ms (2048 samples) and 10ms (441 samples). The low and high frequency bounds are 27.5Hz and 16kHz. we use log-mel input features with a context size of 15 frames and 80 bins as inputs to the CNN. Thus the CNN model takes a binary onset/non-onset decision sequentially for every frame given its context: $\pm$70ms, 15 frames in total. This audio pre-processing configuration is almost the same as in Schl\"{u}ter and B\"{o}ck's work\cite{schluter2014improved} except that 3 input channels with respectively frame sizes 23ms, 46ms and 93ms have been used in their work, whereas only 1 channel with frame size 46.4ms input is used in this research.

\subsection{Deep learning onset detection functions}\label{sec:nn_onset}
In this section, we introduce the neural network setups and training strategies for two experiments. The first experiment aims to find the most efficient network architecture trained separately on two datasets for the onset detection. The second aims to study the inter-dataset onset detection performance by applying the models or feature extractors pre-trained on one dataset to another dataset.
\subsubsection{Searching for the most efficient neural network architecture}\label{sec:architectures}
Following the terminology used in Pons et al.'s work\cite{pons2017score}, we regard a neural network architecture as two parts -- front-end and back-end. According to their work, the front-end is the part of the architecture which processes the input features and maps it into a learned representation. The back-end predicts the output given the learned representation. In this research, we don't restrict the functionality of back-end to prediction. However, we use it as terminology to differentiate from the front-end. We present the front-ends in table \ref{table:frond_ends} and back-ends in table \ref{table:back_ends}. \textbf{Conv} means convolutional layer. \textbf{10x} $\boldsymbol{3\times7}$ means 10 filters of which each convolves on 3 frequency bins and 7 temporal frames. All the Conv layers use ReLU activations. The first Conv layer in the front-end B has 6 different filter shapes. Each Conv layer in back-end C and D follows by a batch normalization layer to accelerate the training\cite{IoffeS15batch}. \textbf{BiLSTMs} means bidirectional RNN layers with LSTM units. In back-end B, both forward and backward layers in BiLSTMs have 30 units with Tanh activations. The activation function type of \textbf{Dense} layer -- ReLU or Sigmoid used in back-end A depends on the architecture.

\begin{table}[ht!]
\centering
\caption{Architecture front-ends}
\label{table:frond_ends}
\begin{tabular}{c|c}
\toprule
Front-end A & Front-end B \\
\midrule
Conv 10x $3{\times}7$ & \makecell{Conv 24x $1{\times}7$, 12x $3{\times}7$, 6x $5{\times}7$\\24x $1{\times}12$, 12x $3{\times}12$, 6x $5{\times}12$} \\
Max-pooling $3{\times}1$ & Max-pooling $5{\times}1$ \\
Conv 20x $3{\times}3$ & Conv 20x $3{\times}3$ \\
Max-pooling $3{\times}1$ & Max-pooling $3{\times}1$\\
Dropout 0.5 & Dropout 0.5\\
\bottomrule

\end{tabular}
\end{table}

\begin{table}[ht!]
\centering
\caption{Architecture back-ends}
\label{table:back_ends}
\begin{tabular}{c|c}
\toprule
Back-end A  & Back-end B    \\
\midrule
Dense 256 units & Flatten \\
Flatten & BiLSTMs 30 units \\
Dropout 0.5 & Dropout 0.5\\

\toprule
Back-end C  & Back-end D    \\
\midrule
Conv 40x $3{\times}3$ & Conv 60x $3{\times}3$ \\
Conv 40x $3{\times}3$ & Conv 60x $3{\times}3$ \\
Conv 40x $3{\times}3$ & Conv 60x $3{\times}3$ \\
Conv 80x $3{\times}3$ & Flatten \\
Conv 80x $3{\times}3$ & Dropout 0.5 \\
Conv 80x $3{\times}3$ & \\
Conv 135x $3{\times}3$ & \\
Flatten & \\
Dropout 0.5 & \\
\bottomrule
\end{tabular}
\end{table}

We present seven architectures which are the combination pipelines of the front-ends and back-ends. All back-ends are connected with a sigmoid unit to output the ODF for the input log-mel contexts.

\noindent\textbf{Baseline}: Front-end A + back-end A with sigmoid activations. This architecture is the same as the one described in Schl\"{u}ter and B\"{o}ck's work\cite{schluter2014improved}.

\noindent\textbf{ReLU dense}: Front-end A + back-end A with ReLU activations. In Schl\"{u}ter and B\"{o}ck's work\cite{schluter2014improved}, using ReLU activations in the back-end A caused a drop in performance when evaluating on B\"{o}ck dataset. However, ReLU activation function has been shown to enable better training of deeper networks because it has several advantages compared with Sigmoid, such as reducing the likelihood of vanishing gradient\cite{glorot2011deep}. We want to (re-)test the performance of ReLU activation on both B\"{o}ck and jingju dataset.

\noindent\textbf{No dense}: Front-end A + Flatten layer. We use this architecture to test the effect of removing the dense layer in the baseline.

\noindent\textbf{Temporal}: Front-end B + back-end A with sigmoid activations. This one is similar to the ``Temporal architecture'' presented in Pons et al.'s work\cite{pons2017score}, and uses various filter shapes in the first convolutional layer. It has been shown experimentally that on a smaller jingju dataset, this architecture outperformed the baseline by effectively learning the temporal onset patterns.

\noindent\textbf{BiLSTMs}: Front-end A with time-distributed Conv layers + back-end B. This one is similar to the CRNNs architectures presented in Vogl et al.'s work\cite{Vogl2017DrumTV}. We use the sequence of the log-mel contexts as the architecture input and we experiment 3 different sequence lengths -- 100, 200 and 400 frames. At the training phase, two consecutive input sequences are overlapped but their starting points are distanced by 10 frames. At the testing phase, the consecutive input sequences are not overlapped. We use this architecture to test the effect of replacing the dense layer in the baseline by RNN layer.

\noindent\textbf{9-layers CNN}: Front-end A + back-end C. We use this architecture to test the performance of deep CNN without using dense layer.

\noindent\textbf{5-layers CNN}: Front-end A + back-end D. As our datasets are relatively small, the above 9-layers CNN could be overfitting. Thus, we test also this shallow architecture with 5 CNN layers.

\begin{table}[ht!]
\centering
\caption{Total numbers of trainable parameters (TNoTP) of each architecture.}
\label{table:parameters}
\resizebox{\columnwidth}{!}{
\begin{tabular}{cccc}
\toprule
Baseline   & ReLU dense   & No dense     & Temporal \\
\midrule
289,273    & 289,273      & 3,161        & 283,687  \\
\toprule
BiLSTMs & 9-layers CNN & 5-layers CNN &          \\
\midrule
278,341    & 288,286      & 81,541       &      \\
\bottomrule
\end{tabular}
}
\end{table}

The TNoTP of each architecture is shown in table \ref{table:parameters}. To keep a fair comparison, we maintain a similar TNoTP between the baseline, ReLU dense, Temporal, BiLSTMs and 9-layers CNN architectures. We reduce the parameter numbers in No dense and 5-layers CNN architectures to explore the model efficiency. Notice that 9-layers and 5-layers CNNs are not fully-convolutional architectures\cite{long2015fully} since we don't perform average pooling to the last Conv layer.

\subsubsection{Inter-dataset transfer learning}\label{sec:transfer_learn}
The most efficient architecture evaluated separately on two datasets (see section \ref{sec:nn_results}) -- 5-layers CNN, is used to conduct the inter-dataset transfer learning experiments. In the context of this research, transfer learning means twofold techniques -- (i) initializing the model by the weights trained on one dataset then re-training the model on another dataset, (ii) using the weights-fixed model pre-trained on one dataset as a feature extractor, combining it with another network and training on another dataset. These transfer learning techniques have been used in image recognition\cite{Razavian2014CNN} and music classification/regression\cite{choi2017transfer} tasks. They have achieved a quite acceptable performance considering that the dataset of the target task is relatively small or the implementation code is not available. We conduct three experiments reciprocally on our two datasets using these two techniques.

\noindent\textbf{Pre-trained}: we directly use the model pre-trained on one dataset to evaluate on another dataset without re-training. This experiment tests whether the pre-trained model is overfitted on the source dataset.

\noindent\textbf{Re-trained}: We first pre-train the model weights on one dataset, then re-train them on another dataset, while keeping the same learning rate. The typical suggestion is to use a smaller learning rate when re-training on the target dataset because the regular learning rate might distort the original weights too quickly\cite{standford2017transfer}. However, we keep using the same learning rate for both train and re-train processes to maintain the consistency of this hyperparameter.

\begin{figure}[ht!]
    \centering
    \includegraphics[width=0.5\textwidth]{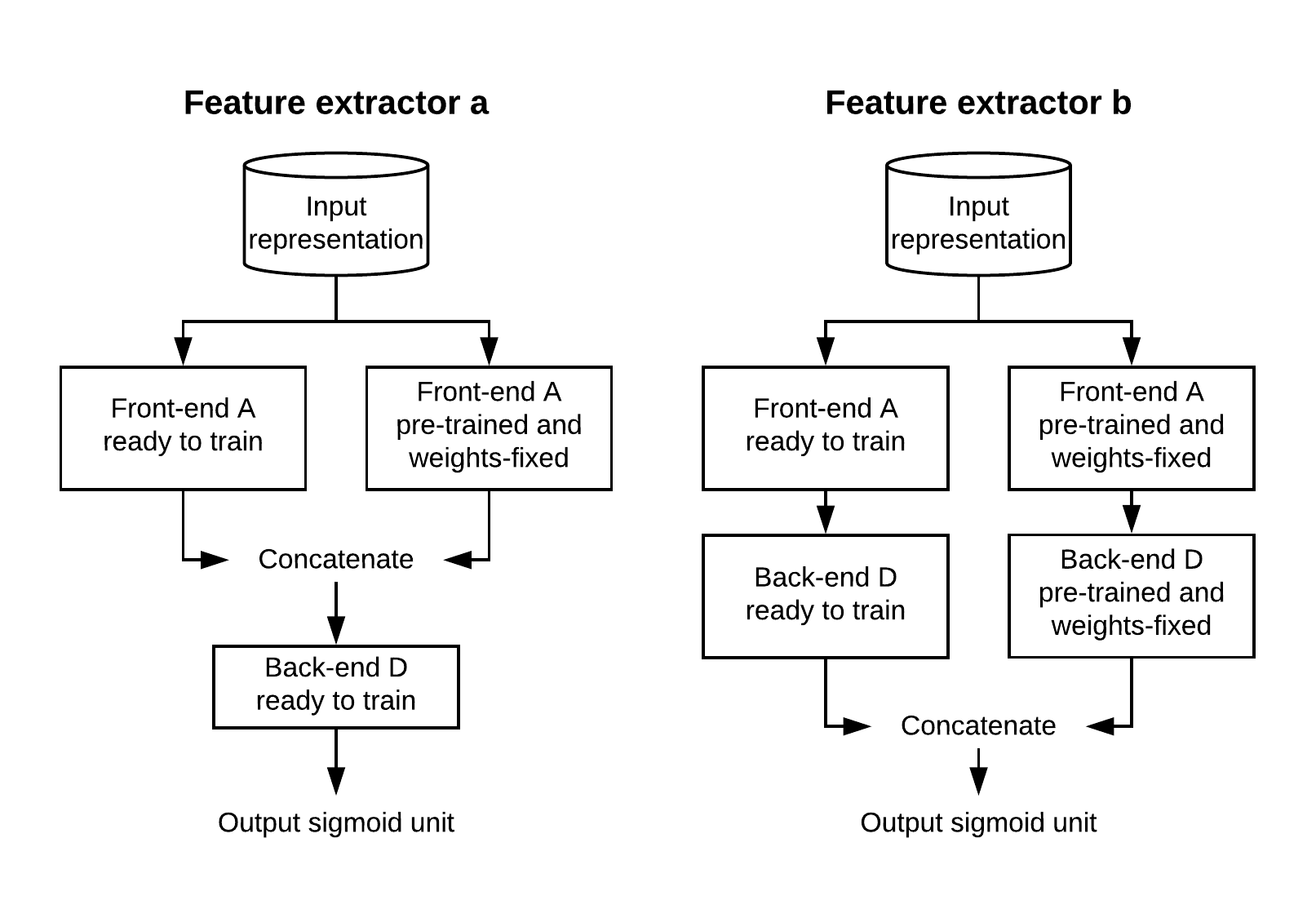}
    \caption{Feature extractor architectures.}
    \label{fig:feature_extractors}
\end{figure}

\noindent\textbf{Feature extractor}: We fix the weights of 5-layers CNN model pre-trained on one dataset, use either (a) front-end A or (b) front-end A + back-end D as the feature extractor, ensemble it into another 5-layers CNN, then train the entire architecture on the target dataset. These two strategies are illustrated in figure \ref{fig:feature_extractors}. This experiment tests whether the feature learned from the source dataset can help learning the target data.

\subsubsection{Model training}\label{sec:model_training}

We use the same target label preparing strategy been described in Schl\"{u}ter and b\"{o}ck's work\cite{schluter2014improved}. The target labels of the training set are prepared according to the ground truth annotations. We set the label of a certain context to 1 if an onset has been annotated for its corresponding frame, otherwise 0. To compensate the human annotation inaccuracy and to augment the positive sample size, we also set the labels of the two neighbor contexts to 1. However, the importance of the neighbor contexts should not be equal to their center context. Thus the sample weights of the neighbor contexts are compensated by being set to 0.25. The labels are used as the training targets in the deep learning models to predict the onset presence.

Binary cross-entropy is used as the loss function. The model parameters are learned with mini-batch training (batch size 256), Adam\cite{KingmaB2014adam} update rule. 10\% training data is separated in a stratified way for early stopping -- if validation loss is not decreasing after 15 epochs. In all experiments, we use \texttt{Keras}\footnote{\url{https://github.com/keras-team/keras}} with \texttt{Tensorflow}\footnote{\url{https://github.com/tensorflow/tensorflow}} backend to train the models.

\subsection{Onset selection}\label{sec:onset_selection}
The ODF output from the model is smoothed by convoluting with a 5 frames Hamming window. Onsets are then selected on the smoothed ODF. Two onset selection methods are evaluated.  The first is a peak picking method which has been used in many MOD works\cite{Bock2012EvaluatingTO,schluter2014improved,Vogl2017DrumTV}. We use the \texttt{OnsetPeakPickingProcessor} module implemented in \textsc{Madmom}\cite{madmom2016} package. Please refer to our code for its detailed parameter setting. Another onset selection method is based on a score-informed HMM\cite{gong2017score,pons2017score}, which has been used to take advantage of the prior syllable duration information of the musical score.

\section{Experimental setup}\label{sec:eval} 
A detected onset is considered as correct if it is not farther
than 25 ms from an unmatched manually annotated onset. We report the peak-picking evaluation results on both B\"{o}ck and jingju datasets. The jingju dataset has the musical score, which allows us to report also the result using score-informed HMM onset selection method.

We use the same splits for 8-fold CV as in \cite{schluter2014improved}. The models are trained on the training sets and tested on their corresponding holdout test sets. The peak-picking results are reported by grid searching the best threshold on the holdout test set. The same evaluation setup has been described in Schl\"{u}ter and B\"{o}ck's work\cite{schluter2014improved}. For the jingju dataset, the pick-peaking results are reported by grid searching the best threshold on the test set, and the score-informed HMM results are evaluated directly on the test set since no optimization is needed.

We report only F1-measure in this paper. 
For jingju dataset, to average out the network random initialization effect, each model is trained 5 times with different random seeds, then the average and standard deviation results are reported. For B\"{o}ck dataset, we train each model only one time because the use of the 8-fold cross validation, where different random initializations are performed for different folds. To measure the statistical significance of the performance improvement or deterioration, we calculate the Welch’s t-test on the 8 folds results for B\"{o}ck dataset and 5 training times results for jingju dataset. We report two tails p-value and reject the null hypothesis if the p-value is smaller than 0.05.

\section{Results and discussion}
In this section, we report and analyze the results for the most efficient architecture searching and the inter-dataset transfer learning experiments. In tables \ref{table:peak_picking_bock_archi} and \ref{table:peak_picking_jingju_archi}, the p-value is calculated by comparing each model results with \textbf{Baseline}. Whereas in tables \ref{table:peak_picking_bock_transfer} and \ref{table:peak_picking_jingju_transfer}, it is calculated by comparing with \textbf{5-layers CNN} results.

\subsection{Searching for the most efficient neural network architecture}\label{sec:nn_results}

Observing table \ref{table:peak_picking_bock_archi} -- the results of B\"{o}ck dataset, no architecture is significantly better or worse than \textbf{Baseline}. \textbf{ReLU dense} achieves equal performance to \textbf{Baseline} (0.990 p-value), which contradicts to the result reported in Schl\"{u}ter and B\"{o}ck's work\cite{schluter2014improved}. Additionally, the training of \textbf{ReLU dense} is significantly faster than \textbf{Baseline} because of the use of ReLU activations. \textbf{BiLSTMs} and \textbf{9-layers CNN} give a slightly worse performance although we would have intended to profit from the RNN or deep CNN layers. The worse performance is due to overfitting, which can be explained by their train and validation loss curves (check them on the Github page\footnote{\url{https://github.com/ronggong/musical-onset-efficient}\label{fn:github}}). \textbf{BiLSTMs 200} (200 means the input sequence length) has a quite low score because we found that its fold 4 model fails drastically in evaluating the holdout test set (F1-measure: 73.25\%), and we believe that this is due to a poor random initialization in training this fold. 

\begin{table}[ht!]
\centering
\caption{B\"{o}ck dataset peak-picking results of different architectures.}
\label{table:peak_picking_bock_archi}
\begin{tabular}{l|cc}
\toprule
 			   & F1-measure & p-value \\
\midrule
Baseline       & 86.67 & -- \\
ReLU dense     & 86.65 & 0.990 \\
No dense       & 85.38 & 0.083 \\
Temporal       & 87.07 & 0.486 \\
BiLSTMs 100 & 85.08 & 0.085 \\
BiLSTMs 200 & 83.04 & 0.077 \\
BiLSTMs 400 & 85.41 & 0.068 \\
9-layers CNN   & 85.78 & 0.246 \\
5-layers CNN   & 86.52 & 0.813 \\
\bottomrule
\end{tabular}
\end{table}

\textbf{Baseline} is different from the state-of-the-art model described in Schl\"{u}ter and B\"{o}ck's work\cite{schluter2014improved} regarding that the latter used (i) a 3 channels input representation, and (ii) a stochastic gradient descent (SGD) optimizer with gradually decreased learning rate and momentum and a fixed number of 300 training epochs. We have also tested (i) the 3 channels input representation, which gave 0.497\% F1-measure improvement (87.16\%), and tested the learning rate configuration (ii), which didn't improve the F1-measure. Thus, we keep using 1 channel representation and adam optimizer in our experiments. Besides, our goal is to find an efficient architecture rather than surpassing the performance of the state-of-the-art.

\begin{table}[ht!]
\centering
\caption{Jingju dataset peak-picking (upper) and score-informed HMM (bottom) results of different architectures.}
\label{table:peak_picking_jingju_archi}
\begin{tabular}{l|cc}
\toprule
               & F1-measure & p-value \\
\midrule
Baseline	   &76.17$\pm$0.77 		 & -- \\
ReLU dense     &76.04$\pm$1.02       & 0.840              \\
No dense       &73.88$\pm$0.44       & 0.002              \\
Temporal       &76.01$\pm$0.61       & 0.749              \\
BiLSTMs 100 &78.24$\pm$0.83       & 0.006              \\
BiLSTMs 200 &77.82$\pm$0.68       & 0.013              \\
BiLSTMs 400 &76.93$\pm$0.68       & 0.178              \\
9-layers CNN   &73.83$\pm$0.92        & 0.005              \\
5-layers CNN   &76.68$\pm$1.04         & 0.457             \\

\toprule
               & F1-measure & p-value \\
\midrule
Baseline	   & 83.23$\pm$0.57			& -- 				\\	
ReLU dense     & 82.49$\pm$0.28        & 0.057              \\
No dense       & 82.19$\pm$0.44         & 0.021              \\
Temporal       & 83.23$\pm$0.57         & 1                  \\
BiLSTMs 100 & 82.99$\pm$0.31         & 0.479              \\
BiLSTMs 200 & 83.29$\pm$0.37         & 0.882              \\
BiLSTMs 400 & 82.47$\pm$0.54        & 0.087              \\
9-layers CNN   & 80.90$\pm$0.67         & 0.001              \\
5-layers CNN   & 83.01$\pm$0.76        & 0.649             \\
\bottomrule
\end{tabular}
\end{table}

Observing table \ref{table:peak_picking_jingju_archi} -- the results of jingju dataset, \textbf{BiLSTMs 100} and \textbf{200} outperform \textbf{Baseline} with peak-picking onset selection method but not with score-informed HMM method. \textbf{9-layers CNN} overfits and significantly performs worse than \textbf{Baseline}, which means this architecture is too ``deep" for jingju dataset (check the Github page\footnotemark[\getrefnumber{fn:github}] for its loss curve). \textbf{Temporal} architecture has the p-value of 1 when evaluating by score-informed HMM method, and we confirm that it is a coincidence after having checked its 5 training times F1-measures. \textbf{No dense} architecture performs significantly worse than \textbf{Baseline}. However, considering its tiny TNoTP -- 3,161, this performance is quite acceptable. The similar case has been reported in Lacoste and Eck's work\cite{Lacoste2007supervised}, where their 1 unit 1 hidden layer architecture achieved a remarkable result (only 4\% F1-measure worse than their best architecture). This means that if the state-of-the-art performance is not required, one can use a quite small and efficient architecture. The score-informed HMM onset selection method outperform the peak-picking by a large margin. Also notice that the score-informed HMM method is able to compensate both good and bad performance of peak-picking, which can be seen by comparing upper and bottom results regarding \textbf{No dense}, \textbf{BiLSTMs 100} and \textbf{200} models.

Finally, we choose \textbf{5-layers CNN} as the most efficient architecture because it performs consistently equivalent to \textbf{Baseline} but only contains 28.3\% TNoTP. Although \textbf{Temporal} architecture performs equally well, it is not selected because its equal TNoTP to \textbf{Baseline} and the complex configuration of its front-end B. \textbf{BiLSTMs} outperforms \textbf{Baseline} on jingju dataset, however, due to its overfitting on B\"{o}ck dataset and slow training, we don't consider it as an efficient architecture.

\subsection{Inter-dataset transfer learning}\label{sec:transfer_results}

Observing table \ref{table:peak_picking_bock_transfer} -- the results of B\"{o}ck dataset, no transfer learning method can achieve a better performance than the \textbf{5-layers CNN} model trained directly on B\"{o}ck dataset. \textbf{Pre-trained} model on jingju dataset fails drastically. \textbf{Re-trained} and \textbf{Feature extractor} strategies also show no effect in leveraging the performance.

\begin{table}[ht!]
\centering
\caption{B\"{o}ck dataset peak-picking results of transfer learning experiments.}
\label{table:peak_picking_bock_transfer}
\begin{tabular}{l|cc}
\toprule
                      & F1-measure & p-value \\
\midrule
5-layers CNN          & 86.52               & --          \\
Pre-trained           & 38.73               & 1.08E-08    \\
Re-trained            & 85.85               & 0.387       \\
Feature extractor a   & 86.31               & 0.797       \\
Feature extractor b   & 86.07               & 0.593       \\
\bottomrule
\end{tabular}
\end{table}

\begin{table}[ht!]
\centering
\caption{Jingju dataset peak-picking (upper) and score-informed HMM (bottom) results of the transfer learning experiments.}
\label{table:peak_picking_jingju_transfer}
\begin{tabular}{l|cc}
\toprule
                      & F1-measure & p-value \\
\midrule
5-layers CNN          & 76.68$\pm$1.04              & --        \\
Pre-trained           & 34.28$\pm$1.44              & 2.26E-10  \\
Re-trained & 77.34$\pm$0.84              & 0.354     \\
Feature extractor a   & 76.46$\pm$0.64              & 0.727     \\
Feature extractor b   & 77.27$\pm$0.40              & 0.339     \\

\toprule
                      & F1-measure & p-value \\
\midrule
5-layers CNN          & 83.01$\pm$0.76              & --         \\
Pre-trained           & 38.02$\pm$1.33              & 5.78E-10   \\
Re-trained			  & 83.02$\pm$1.12              & 0.980      \\
Feature extractor a   & 81.91$\pm$0.72              & 0.070      \\
Feature extractor b   & 83.29$\pm$0.62              & 0.588      \\
\bottomrule
\end{tabular}
\end{table}

Observing table \ref{table:peak_picking_jingju_transfer} -- the results on jingju dataset, no transfer learning method can achieve a significant improvement than the \textbf{5-layers CNN} trained directly on jingju dataset. Again, \textbf{Pre-trained} model on B\"{o}ck dataset fails drastically. The score-informed HMM onset selection method is preferable in case that the musical score is available because it can bring a significant improvement compared with peak-picking method.

We believe the reason of the non-improvement of these transfer learning strategies is that the onset temporal-spectral patterns of two datasets are different -- the onset patterns in B\"{o}ck dataset are mainly the attack transients of instrumental sounds, whereas, the patterns in jingju dataset are the syllable attacks of the consonants, semi-vowels or vowels. Thus, the models pre-trained on the source dataset failed to capture the onset patterns of the target dataset.

\section{Reproducibility}\label{sec:reproducibility}
Experiment code
and pre-trained models used in the experiments are available in Github\footnotemark[\getrefnumber{fn:github}]. Jingju dataset is openly available\footnotemark[\getrefnumber{fn:jingju_dataset}]. B\"{o}ck dataset is available on request. A Jupyter notebook is prepared for showcasing the performance of different network architectures\footnote{\url{https://goo.gl/Y5KAFC}}. 


\section{Conclusions}\label{sec:conclusions}
To confront the challenges posed by the previous MOD research, we experimented seven MOD deep learning architectures on two different datasets, of which a 5 layers CNN architecture was identified as the most efficient one. It achieved the equivalent performance of our implementation of the state-of-the-art, however, only contains 28.3\% of the trainable parameters. Two onset selection methods -- peak-picking and score-informed HMM were compared and the latter exhibited a superior performance by benefiting from the prior information of the musical score. Additionally, the results of our inter-dataset transfer learning experiments showed that when the musical content contained in two datasets are different, the pre-trained model tended to capture only the onset patterns of the source dataset, and failed in predicting the onsets on the target dataset. Thus it is important to provide the model training code to enable re-training the model.

\bibliography{ISMIRtemplate}

\newpage
\section{Reviews}
\subsection{Reviewer 1}
The paper discusses general problems of deep learning research like
reproducibility of training, architecture choice, and hyper parameter
selection.
Current state-of-the-art research in automatic onset detection is used and
compared to eight similar architectures to investigate how model choice
influences performance and generalization capabilities.
For this, two datasets are used: a public available dataset created and
commonly used for onset detection evaluation, and a new dataset consisting of
singing voice only.
Additionally a score informed probabilistic model is used and it is confirmed
that score informed onset detection can improve results.
The work further shows that onset detection models which seem to generalize
well within an dataset focused on instrument onset detection fail to be
applicable on a singing voice only dataset.
The final finding is, that a model with a one-fourth of parameters can perform
similarly well on both datasets.

Review:
The work is structured well, the state-of-the art part is nicely done.
Special attention is put into the evaluation to test for significant
differences.
Although the paper is well understandable there are several points which could
be improved:

1) 
The work does not use the actual state-of-the-art implementation, which is
publicly available as pretrained models.
The reported results in Schl\"{u}ters work \cite{schluter2014improved} are significantly higher on the
B\"{o}ck dataset (90.3\% v.s. 86.7\%).
While for a architecture comparison it is not necessarily mandatory to achieve
state-of-the-art performance, the question is if the smaller models which can
achieve similar performance in the present work, would be tunable to achieve
state-of-the-art performance.

2) 
In the work several problems of current deep learning research are discussed,
like architecture choice, hyper parameter tuning, and reproducibility.
While in this work a good effort is made to make the work reproducible, it is
not clear how the used architectures were selected. 
Also the use of input features, no learn rate scheduling and other
hyper-parameter settings which are different to current state-of-the-art
methods are not sufficiently motivated.

\textcolor{red}{Rong Gong: I have tried with the same input representation (3 channels log-mel using Madmom package), network architecture, and gradually decreased learning rate with momentum, a fixed number of 300 training epochs. The 3 channels representation improved the baseline by 0.497\%, and the learning rate scheduling didn't bring improvement.}

3)
In the introduction of the work, several papers are cited when motivating the
need for onset detection.
Most of these paper use, however, deep learning end-to-end methods, thus not
requiring any onset detection at all.
Onset detection is surely an important task, but the choice of works to
motivate it in this works seems odd.

\textcolor{red}{Rong Gong: In the introduction section, beat tracking \cite{Bock2014AMA} and tempo estimation \cite{bock2015accurate} papers use multiple recurrent neural networks to estimate the beat activation functions (beat onsets); drum transcription paper \cite{Vogl2017DrumTV} uses neural networks to generation drum activation functions (drum onsets); note transcription paper \cite{bock2012polyphonic} uses recurrent neural networks to estimate the piano note onset and pitch. Maybe onset detection is not the prerequisite step, but it should be the intermediate step.}

4)
In the shortcomings section, generalization and model capacity problems are
discussed.
While this is my personal opinion, I think it is safe to state that authors
usually make sure their experimental setup counteracts overfitting as well as
the datasets allow.
In case of the mentioned drum transcription experiments e.g., the used datasets
are split using the different sounding drums and drummers, and three-fold
cross-validation is performed.
Overfitting to drum sounds of the training set by means of too much model
capacity, as is mentioned to be a danger, is a real problem but in this kind of
setup, it should show by producing lower performance on the cross-validation
evaluation.

\textcolor{red}{Rong Gong: I removed the drum transcription part in this section.}

5)
It is claimed that the pre-trained models are not sufficient because retraining
is necessary for other datasets.
This claim could easily be given more basis by performing an evaluation using
the pretrained models on the new datsets. 

6) 
It is stated that in Schl\"{u}ters work [30] ReLU in the dense output layers
decrease performance, which can not be reproduced.
The performance drop in \cite{schluter2014improved} is from 90.3 to 89.6\% -- both higher than the
achieved values in the present work. 
The performance also drops for the present work, from 86.67 to 86.65\%. 
While no significance testing was performed in Schl\"{u}ters work, I would assume
both are not significant.

\textcolor{red}{Rong Gong: yes, they are both not significant probably.}

7)
It was surprising, that current state-of-the-art onset detection methods, which
perform quite well in real world evaluation scenarios, like MIREX, fail that
dramatic on a singing voice dataset.
A more detailed investigation why this is the case would have been interesting.
I can only suspect that pure singing voice onset detection is very different
from instrument onset detection.

\subsection{Reviewer 2}

This paper investigates several (similar) neural network (NN) architectures for the task of onset detection. The authors identified the reproducibility of existing works to be a major difficulty because most papers do not provide code for training NN models. But having training code available is important, since pre-trained models do often perform much worse on other kinds of data. The paper shows this with a nice generalisation/transfer learning experiment by comparing the performance on two very different datasets, one of them to be released together with this paper.

However, when proposing their own NN model for onset detection (similar to existing state of the art, but fewer parameters) the authors make some mistakes which question the whole experiments in the paper. As baseline for all further experiments, they use a re-implementation of [30], but fail to achieve similar results (performance: 86.7\% vs. 90.3\% F-measure). Instead of explaining the differences or showing that this difference is statistically not significant (very unlikely), they use this low-performance baseline for all further statistical tests and conclude that their final model performs statistically on par with state of the art. 

\textcolor{red}{Rong Gong: I didn't claim that the difference between 86.7\% and 90.3\% F-measure is not significant. What I said is that, for B\"{o}ck's dataset, the re-implementation of the state-of-the-art (86.7\% F-measure) is not significantly different from other architectures.}

The code and dataset supplementing this paper are unstructured and not really user-friendly or reusable. But this must be expected from a paper with such a strong focus on reproducibility and reusability of the paper.

All in all, this paper can be strengthened a lot by addressing the issues raised by the reviewers. Unfortunately I can't give the authors the benefit of the doubt to address all points in time for the final version, but I strongly encourage the authors to resubmit it to another conference or to ISMIR next year.

(Own) Review of paper \#76
--------------------------

This paper describes a study on various deep neural network architectures for onset detection. It puts its main focus on reproducibility of experiments and proposes a network similar to existing state-of-the-art which reduces the number of trainable parameters to 28\%, while achieving similar results. Performance is tested on 2 datasets, one containing mostly mixed signals and the other Jingju (Bejing opera) singings and it is shown that pre-trained models on one set don't perform well on the other set, but a lack of training code for existing deep learning was identified.

While this paper adds some contributions, namely i) providing source code for training neural network models and ii) investigating inter-dataset generalisation and transfer-learning capabilities, it has a number of flaws:

- it is not able to reproduce the state of the art in onset detection (\cite{schluter2014improved}) nor does it explain the differences observed

\textcolor{red}{Rong Gong: This should be done in the future or the author of the paper \cite{schluter2014improved} should have made the code available for the reproducibility of his work.}

- puts too much focus on parts not relevant towards the main goal (providing an efficient and reproducible deep learning model for onset detection). 

- highly relevant information and motivation why this is needed in the first place is missing

- investigates a vanishing problem, since the task of onset detection is taken over by systems learned end-to-end

\textcolor{red}{Rong Gong: Wow, I am so sorry to hear that, but the work has already been done...}

Thus the paper would benefit from a shift in focus, paper space could be better spent on more important aspects. I can only advise the authors to carefully address these points of criticism. The following remarks on individual sections should help to make it a better paper.

Abstract
--------

The claim "The most efficient one achieves equivalent performance to our implementation of the state-of-the-art architecture" seems to be a bit odd, since the reported performance is considerably worse than the original publication (86.5\% vs 90.3\% F-measure on the same dataset). The authors don't explain where this difference might stem from nor question their training procedure.

Furthermore, I am not sure that "(iii) ignoring the network capability when comparing different architectures" is a shortcoming per se when comparing different network architectures. Different layer/network types have totally different number of parameters and are able to perform equally well.

Introduction
------------

The authors state that "onset detection ... is a prerequisite step for many MIR tasks", however most references listed are counter-examples for this claim. E.g. [7,8,9,32] use all end-to-end learning, i.e. these methods explicitly skip a dedicated onset detection stage but rather learn the features relevant for the task directly from audio (or some low lever representations such as spectrograms).

\textcolor{red}{Rong Gong: please check my response for the 3) point of the reviewer 1.}

The authors did perform some generalisation and transfer-learning tests, but background information on why predicting onsets with a model trained on (completely) different music signals fails, is missing. In my opinion this section lacks insight on the different features of various kinds of music, e.g. the problem of vibrato in opera singing. It should be stated clearly what the authors think a good onset detection algorithm should be able to do and analyse (not necessarily in this section) how existing methods fail to do so.

\textcolor{red}{Rong Gong: this is a very good point, thanks!}

The sentence "The advantage of these methods is no training data needed" is wrong in my opinion. Although the systems are not trained, they still have a number of adjustable parameters which need to be tuned. Thus training data is needed.

Although it is correct that the total number of learnable parameters should be considered when comparing different network architectures, and networks with more parameters tend to better memorise the training data, usually the networks are designed in such a way that they do not overfit to the training data. It is quite possible that a certain architecture with less parameters overfits earlier than another one with more parameters. Thus bigger networks do not necessarily lead to more overfitting.

It is not clear to me how "We experiment two onset selection methods and shows the preferability of using the score-informed method if the musical score is available as a side information (section 5.1)" is a contribution of this work, since this has already be done in [20,27].

Dataset
-------

Although providing download links to the Jingju dataset (on the supplementary page) is highly appreciated and definitely a step in the right direction, a quick investigation on the usefulness of this dataset revealed that it is not really usable in the state it is provided. The archive contains 116 folders with 308 audio files and more than 1.5k files in total. It is by no means clear or stated which files are used, so it is impossible to compare the results of the proposed network architecture/model with existing other works on the same dataset. At least a list with the files used for training/testing and human and machine readable annotations with onset times in seconds from the beginning of the audio files should (must) be made available.

\textcolor{red}{Rong Gong: the list the file names of the training/testing are provided.}

It would be furthermore great to contact the author of the other set to make it freely available as well.

\textcolor{red}{Rong Gong: I have contacted them, they only provide this dataset on request.}

A few more words about the datasets used would make the paper more self-contained, especially regarding the Bejing opera, since a lot of readers are not very familiar with this kind of music an its peculiarities. E.g. what are the phrases mentioned in Table 1?

Method
------

It is more or less clear what "We use a log-mel context as the CNN model input, where the context size is 80×15 (bins×frames)" might mean, but it is a very unfortunate wording. Try rephrasing as "we use log-mel input features with a context size of 15 frames as inputs to the CNN". The wording 'log-mel context' is used in several other places as well, not only here. As a side note, the notation of frames x bins seems to be used way more often than bins x frames. 

The section about target label preparation (3.2) could be merged with model training (3.3.3) since it relevant only for training. Also consider moving 3.3.3 before 3.3.2.

In literature bidirectional LSTM RNNs are usually abbreviated BLSTM (not Bidi LSTM).

Front-end B is hard to understand how filters of different shapes are combined in a single layer. Since this is not a very common setup, a few sentences would help a lot to understand this approach much better.

For the "No dense: Front-end A, no back-end." setting: isn't there a Flatten layer missing somewhere, or how is the output of shape 8x7 fed to the sigmoid classification layer? Please explain if no Flatten layer is used.

\textcolor{red}{Rong Gong: Good point! There should have a Flatten layer there.}

Regarding the inclusion of the second HMM-based score informed onset selection mechanism I am not sure if this is relevant for this paper at all. First of all, this is beyond the main focus of this paper (Title: Towards an efficient and reproducible deep learning model for musical onset detection), and the differences between this method and the simpler peak-picking based one are very consistent for both the search for the best network architecture (Table 6) and transfer-learning (Table 8). In my opinion, the paper would gain if this part is removed and more space is spent on the motivation and explanation on why there is a need for a reproducible and accessible model training procedure.

Experimental Setup
------------------

"The peak-picking results are reported by grid searching the best threshold on the holdout test set." This is exactly what never should be done. Never ever! Either the training set or (if available) the validation set should be used for this, otherwise the supposed to be independent test set is not independent any more.

\textcolor{red}{Rong Gong: Indeed, strictly speaking, this is wrong to use the test set for the grid search. However, this is how the thing has been done in the state-of-the-art paper \cite{schluter2014improved}. I have to maintain this setup to make a fair comparison with their results.}

It is not clear, why for the Jingju dataset 5 training runs are performed, whereas each fold of the B\"{o}ck set is trained only once ("we train each model only one time because the use of the 8-fold cross validation"). Usually training should be fast enough (the dataset is much smaller and network training converges quite quickly) to perform several training runs and report mean and std.dev. values of the results. This would probably also have solved the issue reported with a training run on a specific fold on the B\"{o}ck set ("has a quite low score because we found that its fold 4 model fails drastically in evaluating the holdout test set"). If not, this is a strong indicator that the initialisation or learn rate chosen is not adequate for the given setting.

Results and discussion
----------------------

The sentence "The worse performance is due to overfitting, which can be explained by their train and validation loss curves" is purely speculative since the architectures in question (BLSTM and 9-layer CNN) have not been trained with fewer parameters (in order to prevent this assumed overfitting) to back up this claim. The loss curves alone do not justify this claim.

When establishing the baseline system the authors use a slightly different system than the one proposed in \cite{schluter2014improved}. While the main difference (single FFT instead of 3 parallel ones) is explained and the impact is tested (-0.5\% F-measure), no investigation is performed on why the established baseline performs much worse (87.2\% vs 90.3\%) than the original implementation. Assuming that Schl\"{u}ter et al. did not invent these performance numbers, there must be a crucial difference in the training procedure. It is weird to read that "However, we keep [...] and adam optimizer in our experiments because our goal is to find an efficient architecture instead of surpassing the performance of the state-of-the-art.", given the paper's claim to find an "efficient and reproducible deep learning model for musical onset detection" and provide source code to train own models on any dataset. How valuable is such a code if it is not able to re-produce the state of the art although all details are given?

\textcolor{red}{Rong Gong: I think the value of the code is to prove that the state-of-the-art is not reproducible by re-implementing the details provided in the paper \cite{schluter2014improved}.}

As mentioned earlier, it is expected that the HMM-based onset detection mechanism works better than the simple peak-picking based one. This was shown in other works and is not of great importance for this paper. Both Table 6 and 8 could be reduced in size by half if the HMM results are omitted.

This space could be used to include results (and an analysis thereof!) obtained with pre-trained models of existing works (e.g. [30] since it performs better than the re-implementation) or non deep learning methods such as [10,11] developed specifically for music signals with vibratos.

Conclusions
-----------

While the proposed network architecture "achieved the equivalent performance of our implementation of the state-of-the-art" it is still not clear where the difference to the original implementation stems from. This claim in its current form is misleading! The better performance of the HMM onset detection was expected and shown by other works.

General remarks
---------------

Table captions should be below the table. Independent from the positions, the captions itself are by no means self-contained. The main information is missing, e.g. Table 6 investigates different network architectures and compares their performance on the Jingju dataset.

\end{document}